\documentclass[reprint]{elsarticle}

\usepackage[utf8]{inputenc}
\usepackage[T1]{fontenc}
\usepackage{graphicx}
\usepackage{dcolumn}
\usepackage{amsmath}
\usepackage{amssymb}
\usepackage{array}
\usepackage{multirow}
\usepackage{calc}
\usepackage{color}
\usepackage{graphicx}
\usepackage{float}
\usepackage{bm}
\usepackage{lscape}
\usepackage{lineno,hyperref}
\modulolinenumbers[5]

\newcommand{\icm}{\ensuremath{~\textrm{cm}^{-1}}}% % cm-1

\newcommand{\urusi}{URu$_2$Si$_2$}
\newcommand{\degree}{\ensuremath{^\circ}}

\definecolor{blue}{RGB}{153,205,255}
\definecolor{blue_light}{RGB}{204,229,255}
\definecolor{red}{RGB}{255,153,153}

\journal{Physica B: Condensed Matter}

\begin{document}

\begin{frontmatter}

\title{Raman Active High Energy Excitations in URu$_2$Si$_2$}
%\tnotetext[mytitlenote]{Fully documented templates are available in the elsarticle package on \href{http://www.ctan.org/tex-archive/macros/latex/contrib/elsarticle}{CTAN}.}

%%% Group authors per affiliation:
%\author{Elsevier\fnref{myfootnote}}
%\address{Radarweg 29, Amsterdam}
%\fntext[myfootnote]{Since 1880.}

%% or include affiliations in footnotes:
\author[MPQ,HFML]{Jonathan Buhot}
\author[MPQ]{Yann Gallais}
\author[MPQ]{Maximilien Cazayous}
\author[MPQ]{Alain Sacuto}
\author[INP]{Przemys\l{}aw Piekarz}
\author[CEA]{Gérard Lapertot}
\author[CEA,IMR]{Dai Aoki}
\author[MPQ]{Marie-Aude Méasson\corref{mycorrespondingauthor}}
\cortext[mycorrespondingauthor]{Corresponding author: marie-aude.measson@univ-paris-diderot.fr}

\address[MPQ]{Laboratoire Mat\'eriaux et Ph\'enom\`enes Quantiques, UMR 7162 CNRS, Universit\'e Paris Diderot, B$\hat{a}$t. Condorcet 75205 Paris Cedex 13, France.}

\address[HFML]{High Field Magnet Laboratory, Institute for Molecules and Materials, Radboud University, Toernooiveld 7, 6525 ED Nijmegen, The Netherlands.}

\address[INP]{Institute of Nuclear Physics, Polish Academy of Sciences, 31-342 Krak\`ow, Poland.}

\address[CEA]{Université Grenoble Alpes, INAC-SPSMS, F-38000 Grenoble, France \\ CEA, INAC-SPSMS, F-38000 Grenoble, France.}

\address[IMR]{Institute for Materials Research, Tohoku University, Oarai, Ibaraki 311-1313, Japan.}

\begin{abstract}
We have performed Raman scattering measurements on \urusi~single crystals on a large energy range up to $\sim$~1300~cm$^{-1}$ and in all the Raman active symmetries as a function of temperature down to 15 K. A large excitation, active only in the E$_{g}$ symmetry, is reported. It has been assigned to a crystal electric field excitation on the Uranium site. We discuss how this constrains the crystal electric field scheme of the Uranium ions. Furthermore, three excitations in the A$_{1g}$ symmetry are observed. They have been  associated to double Raman phonon processes consistently with \textit{ab initio} calculations of the phonons dispersion. 
\end{abstract}


\end{frontmatter}

\linenumbers

\section{Introduction}

Despite thirty years of intense experimental and theoretical research, the microscopic nature of the low temperature phase ($T<17.5$ K) of the heavy fermion compound \urusi~remains unknown \cite{mydosh_colloquium:_2011, mydosh_hidden_2014}. This so-called Hidden Order (HO)  phase emerges in the context of a strong mixing between $5f$ electrons of Uranium atoms and the conduction electrons, which makes its identification particularly difficult. Actually, the valence state of the Uranium ions, configurations U$^{3+}$ ($5f^3$ configuration) and U$^{4+}$ (5$f^2$ configuration) or intermediate, is not clearly established and valence fluctuations have been evoked \cite{fujimori_electronic_2012,hassinger_skutterudite_2008,booth_probing_2016}.
Nevertheless many interesting theories have been proposed to explain the nature of HO, among which multipolar orders from quadrupolar to dotriacontapolar \cite{kusunose_hidden_2011-1, ressouche_hidden_2012, haule_arrested_2009, ikeda_emergent_2012, rau_hidden_2012, suzuki_multipole_2014}, local currents \cite{chandra_hidden_2002, fujimoto_spin_2011}, unconventional density wave \cite{riseborough_phase_2012, das_imprints_2014} modulated spin liquid \cite{pepin_modulated_2011, thomas_three-dimensional_2013}, dynamical symmetry breaking \cite{elgazzar_hidden_2009} and hastatic order \cite{chandra_hastatic_2015-1} for the most recent ones. Although many theories are based on and proposed crystal field schemes (at least the ground state level), all attempts to measure directly the crystal electric field (CEF) excitations in URu$_2$Si$_2$ in the paramagnetic state have failed. Only recently, a low energy excitation (1.7~meV) observed by Raman scattering spectroscopy \cite{buhot_symmetry_2014-1, kung_chirality_2015} has been suggested as related to a CEF excitation.

Raman spectroscopy is a powerful tool to address the CEF excitations \cite{devereaux_inelastic_2007, cardona_light_2000, ogita_crystal_2009}, particularly thanks to the selection rules which provide symmetry dependence of each excitations. Here, using electronic Raman spectroscopy, in addition to the usual phonon modes \cite{buhot_lattice_2015}, we report the existence of four high energy excitations in URu$_2$Si$_2$. Three of them, seen the A$_{1g}$ symmetry, are assigned to double phonon processes while a broader one, seen only in the  E$_{g}$ symmetry, is associated to a crystal field excitation.

\section{Methods}

Polarized Raman scattering has been performed in quasi-backscattering geometry with the 532 nm line of a solid state laser. We have used a closed-cycle $^{4}$He cryostat with sample in high vacuum (10$^{-6}$~mbar). By comparing Stokes and anti-Stokes Raman spectra and via the evolution of phonon frequencies with incident laser power, we have estimated the laser heating of the samples at +1~K/mW. The scattered light has been filtered by RazorEdge filter and analyzed by a Jobin Yvon T64000 spectrometer in a simple grating configuration. All spectra are corrected by the Bose factor.
 URu$_2$Si$_2$ single crystals were grown by the Czochralski method using a tetra-arc furnace \cite{aoki_field_2010}. We have used the same samples, polished or freshly cleaved, as in our previous studies \cite{buhot_raman_2013, buhot_lattice_2015, buhot_symmetry_2014-1}. 
By combining different incident and scattered light polarizations and sample geometries,
we have measured the A$_{1g}$, A$_{2g}$, B$_{1g}$, B$_{2g}$ and E$_{g}$ symmetries (Mulliken-Herzberg notation) of the D$_{4h}$ point group (space group n\degree 139) \cite{hayes_scattering_2004} corresponding respectively to $\Gamma_1^+$, $\Gamma_2^+$, $\Gamma_3^+$, $\Gamma_4^+$ and $\Gamma_5^+$ in  Bethe’s notation. For both notations, "g" (gerade) and "+" (even) designate the respect of the space inversion.

\section{Results and discussion}

\begin{figure}[htpb]
\centering
\includegraphics[width=1\linewidth]{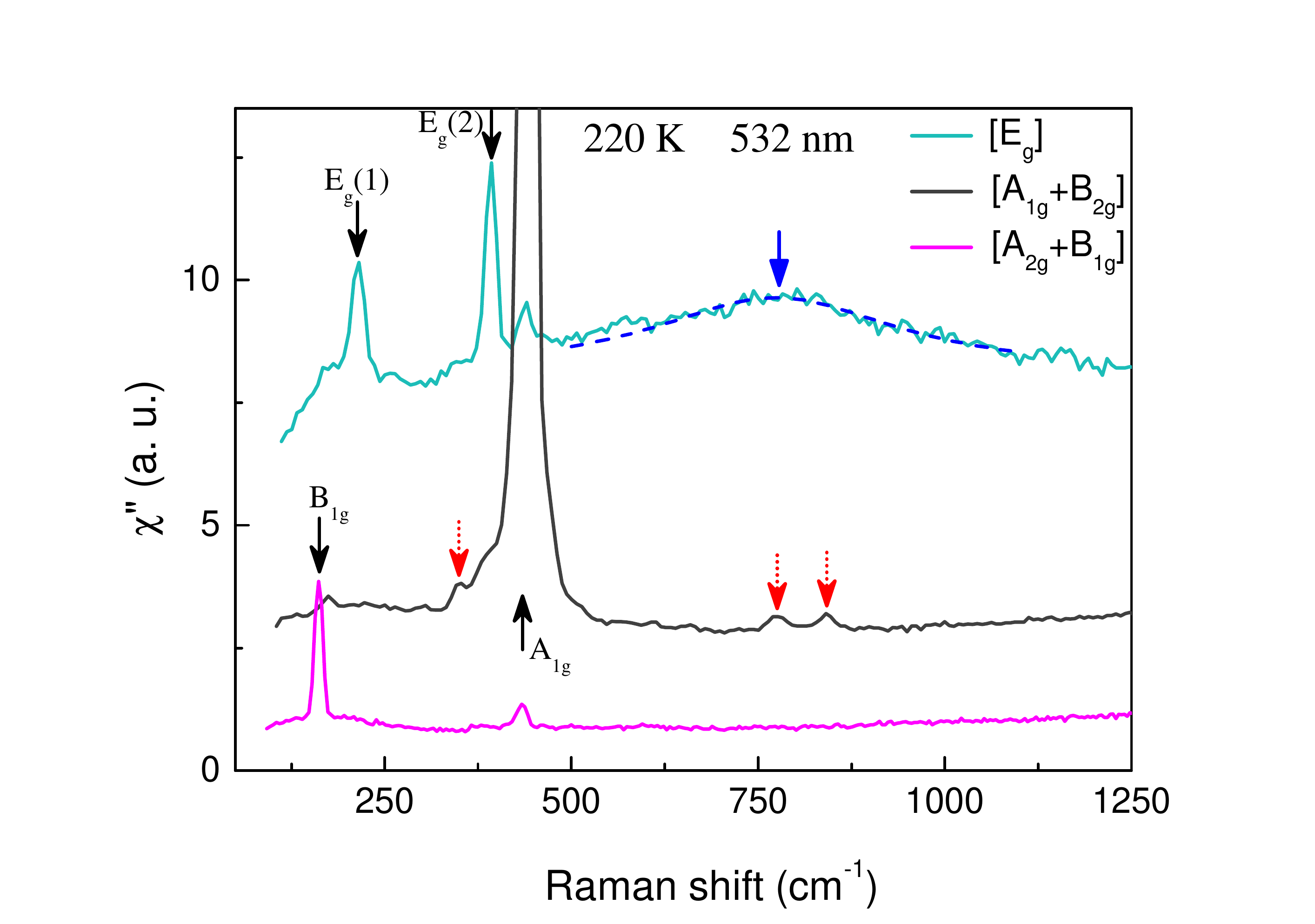}
\caption{Raman susceptibility in a large energy range in [E$_{g}$], [A$_{1g}$ + B$_{2g}$] and [A$_{2g}$ + B$_{1g}$] symmetries at 220 K. Black arrows point to the simple-process phonon modes in each symmetry. Four other peaks are observed: three in the [A$_{1g}$] symmetry (red dotted arrow) assigned to double-process phonon modes, and one in the [E$_{g}$] symmetry  (blue straight arrow). It is attributed to a CEF excitation on the Uranium ions. The blue dashed line corresponds to a Lorentzian fit of this  broad E$_{g}$ excitation.}
\label{fig1}
\end{figure}

\begin{figure}[htpb]
\centering
\includegraphics[width=1\linewidth]{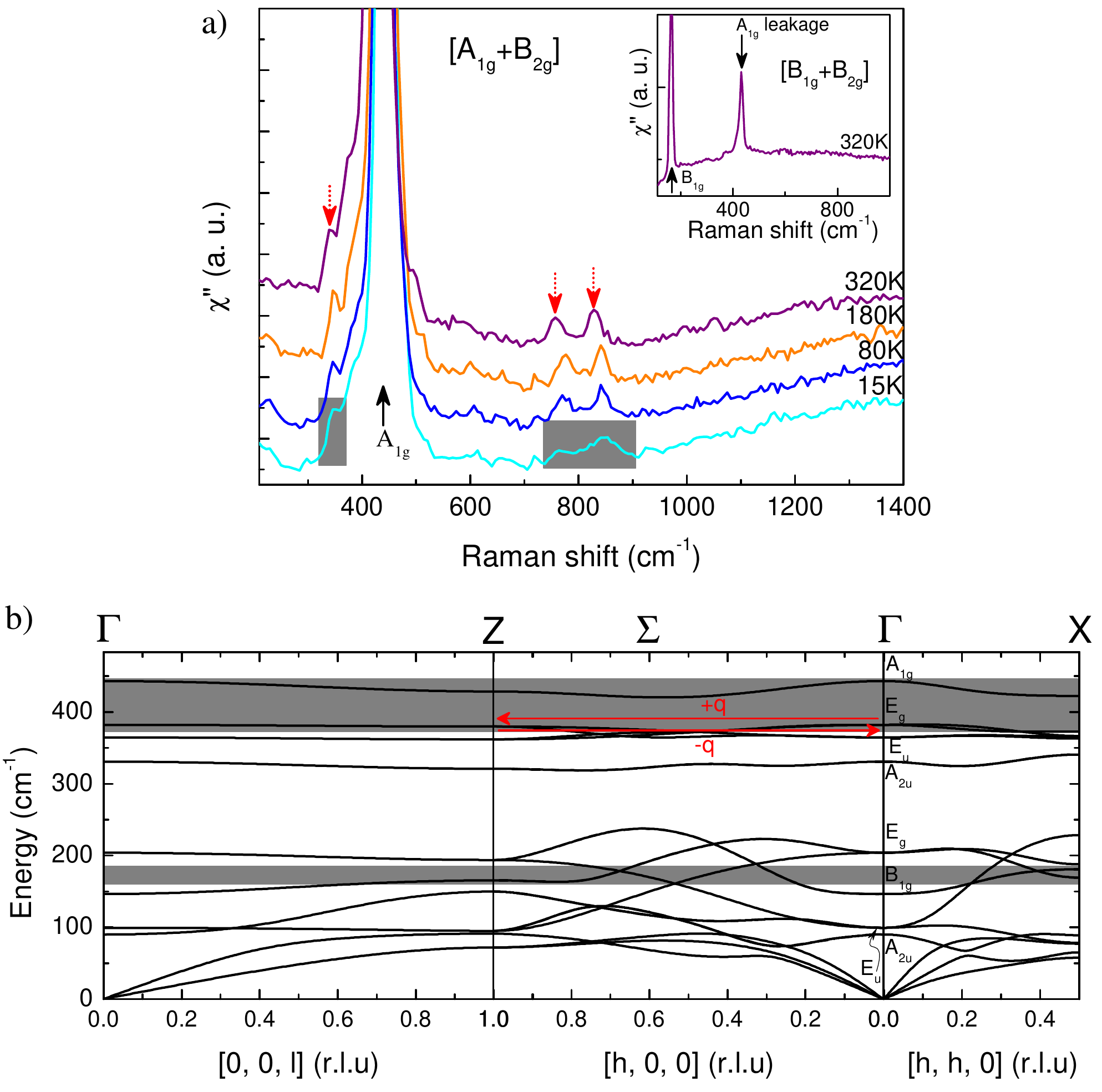}
\caption{a) Raman susceptibility  in the [A$_{1g}$ + B$_{2g}$] symmetry at various temperatures. Spectra have been shifted for clarity. Inset : Raman spectrum in the [B$_{1g}$ + B$_{2g}$] symmetry. Simple phonon modes are denoted by black arrow. Red dot arrows point the three double phonon peaks which are of pure A$_{1g}$ symmetry. The grey areas indicate the energy range of the double phonon processes. b) {\it Ab initio} calculations of the phonon dispersion curves along $\Gamma Z$, $\Gamma \Sigma Z$ and $\Gamma X$ directions at 0~K \cite{buhot_lattice_2015}. The electronic structure was optimized including the spin-orbit coupling within the VASP program \cite{kresse_efficient_1996-1} and the phonon dispersion relations were obtained using the direct method with atomic displacements $u=0.06$ \AA\ \cite{parlinski_first-principles_1997}. The two ($\pm\vec{q}$) red arrows along the [h,0,0] direction depict one possible Raman active double phonon process on the "E$_{g}$" branch. The energy range of the grey areas shown in a) is reported on the dispersion curves.}
\label{fig2}
\end{figure}

Figure \ref{fig1} shows Raman spectra of \urusi~up to 1250\icm~at 220 K for different symmetries. The intense sharp peaks observed at $\sim$ 161\icm, $\sim$ 214\icm, $\sim$ 391\icm~and $\sim$ 433\icm~which correspond to the B$_{1g}$, E$_{g}$(1), E$_{g}$(2) and A$_{1g}$ phonon modes, respectively (See black arrows Fig. \ref{fig1}). Energies of all these phonon modes are consistent with our previous comprehensive study \cite{buhot_raman_2013, buhot_lattice_2015}. In [E$_{g}$] and [A$_{2g}$ + B$_{1g}$] symmetries, a weak leakage of the A$_{1g}$ phonon mode is seen due to a small crystal misalignment. Three new excitations in the [A$_{1g}$ + B$_{2g}$] symmetry and a broad excitation only in [E$_{g}$] symmetry are reported (See respectively, red dot and straight blue arrows on the Figure \ref{fig1}). Very likely, the peaks in the [A$_{1g}$ + B$_{2g}$] symmetry correspond to double Raman phonon processes with a pure A$_{1g}$ symmetry while the peak in E$_{g}$ symmetry originates from CEF excitation. The following sections describe further these assignments.

\subsection{Double phonon processes in the A$_{1g}$ symmetry}

Figure \ref{fig2} a) presents large energy scale Raman spectra in the [A$_{1g}$ + B$_{2g}$] symmetry for several temperatures down to 15 K. At 320 K, the three new broad peaks are seen respectively at 344, 758 and 829~cm$^{-1}$ with a linewidth of $\sim$ 30~cm$^{-1}$ (See red dotted arrows in Fig. \ref{fig2} a)). A slight hardening is observed with decreasing temperature. These new excitations are of pure A$_{1g}$ symmetry since they are not observed in the B$_{1g}$ + B$_{2g}$ symmetry (See inset in Fig. \ref{fig2} a)). 

First, we rule out single phonon processes because all the Raman active phonon modes have been already observed and assigned. Most probably, they are signatures of double phonon processes. Indeed, their energy ranges are in good agreement with \textit{ab initio} calculations of the phonon dispersion \cite{buhot_lattice_2015}. As shown in the Figure \ref{fig2} b), phonon excitations due to a double process (with a total transfered wave-vector that must be zero, i.e. $\vec{Q} =\pm\vec{q} = 0$) can be  Raman active along various directions (such as [h,0,0] shown in Fig\ref{fig2} b), their energy ranges (grey areas) match the Raman shift of the three excitations. They can originate from pure or mixed double processes on the  B$_{1g}$, E$_g$ or A$_{1g}$ branches.

\subsection{Crystal electric field excitation in the E$_g$ symmetry}

\par
\begin{figure}[htbp]
\centering
\includegraphics[width=1\linewidth]{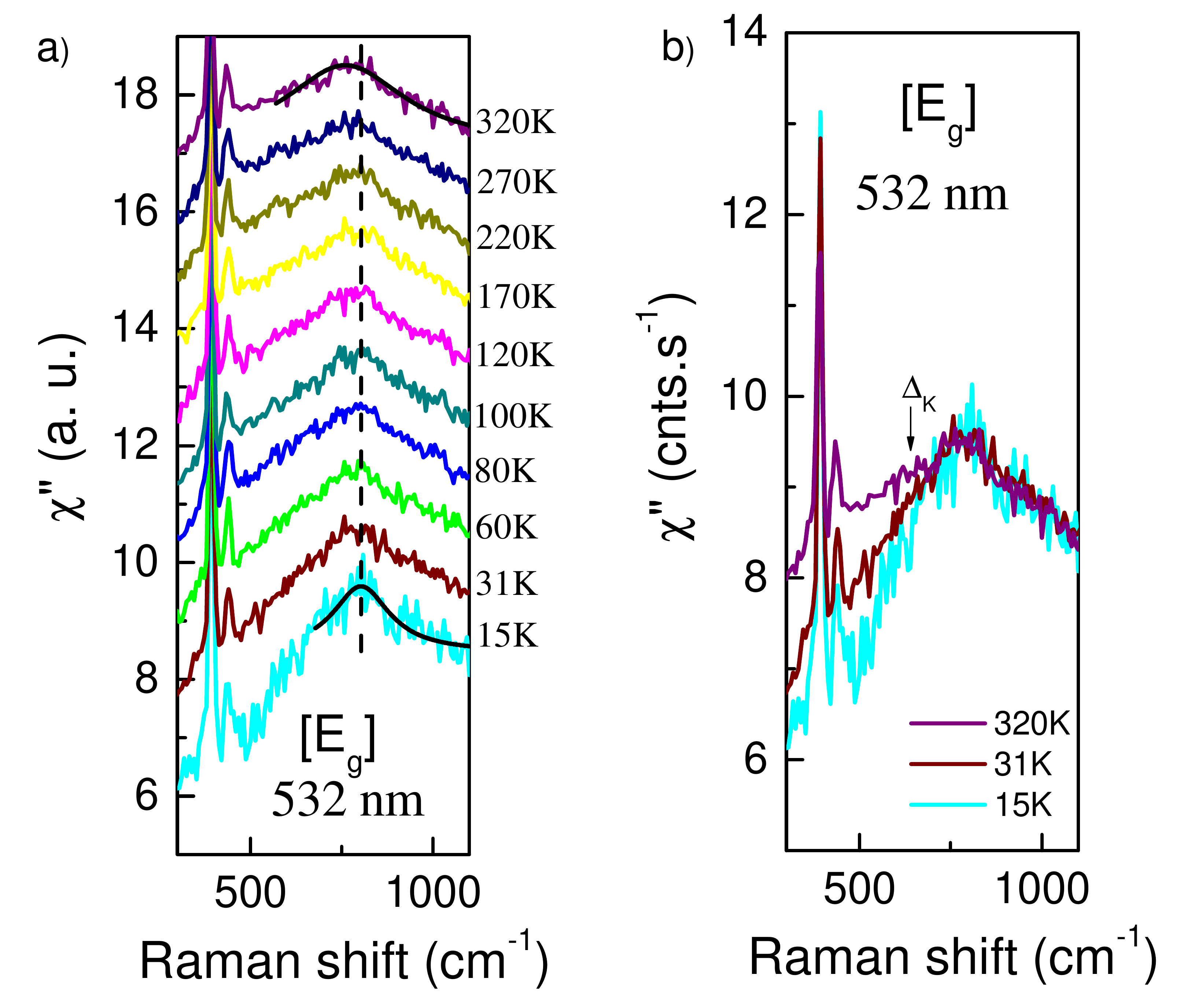}
\caption{a) Raman susceptibility in [E$_g$] symmetry at various temperature showing a possible crystal field excitation. The spectra have been shifted vertically. Black curves at 300 K and 15 K are Lorentzian fits. The E$_g$ peak slightly shifts to high energy upon cooling down. b) Raman susceptibility in [E$_g$] symmetry at 320 K, 31 K and 15 K. $\Delta_K$ denotes the energy of closure of the Kondo gap \cite{buhot__2016}.}
\label{fig3}
\end{figure}

A very broad excitation (FWHM = $\sim 400$ \ensuremath{~\textrm{cm}^{-1}}) is measured at $\sim 760$ \ensuremath{~\textrm{cm}^{-1}} at room temperature (See Fig.\ref{fig3}). As mentioned previously, this broad excitation is observed only in the E$_g$ symmetry (See Fig.\ref{fig1}). Upon cooling the sample, the energy of this large excitation slightly increases up to $\sim 800$ \ensuremath{~\textrm{cm}^{-1}} at 15 K and its width decreases down to $\sim 190$ \ensuremath{~\textrm{cm}^{-1}}. This E$_g$ excitation, much larger than the three other excitations seen in the A$_{1g}$ symmetry, is unlikely to be due to a double phonon process. 
We also rule out the possibility of it coming from a fluorescence effect. Indeed, the fluorescence process is independent of the orientation of the polarization of the incident light. This broad excitation  vanishes when the orientation of the polarization of the incident light is modified. For instance, we have observed the broad E$_g$ excitation in the $x(\mathbf{y},z)\bar{x}$ (E$_g$ probed) configuration\footnote{Porto's notation: $\boldsymbol{k_i}(\boldsymbol{e_i},\boldsymbol{e_d})\boldsymbol{k_d}$} where x, y, z correspond to the crystallographic axis, whereas it disappears when the polarization of the incident light is turned in the $x(\mathbf{z},z)\bar{x}$ (A$_{1g}$ probed) configuration.
An excitation at the same energy has been extracted by inelastic neutron scattering experiment by Park et al. \cite{park_high-energy_2002} where ThRu$_2$Si$_2$ was used as a phonon background material. Indeed, authors report three excitations, one is on line with our observation with energy of $798\pm8$ \ensuremath{~\textrm{cm}^{-1}} and FWHM = $290\pm16$ \ensuremath{~\textrm{cm}^{-1}} at 22 K. They claimed this excitation to be a  crystal electric field one. Consequently, the broad peak in the E$_g$ symmetry observed directly by Raman scattering is most probably a crystal electric field (CEF) excitation ($\Gamma_i\longrightarrow\Gamma_f$, where $\Gamma_i$ has to be a low energy state) from localized $5f$ electrons of Uranium ions. Measuring a CEF excitation with such large width is certainly inherent to the dual character localized/itinerant of the \textit{5f} electrons.

Park et al. claim that their result is consistent with the CEF scheme proposed by Santini and Amoretti’s \cite{santini_crystal_1994}, i.e. a transition $\Gamma_3\longrightarrow\Gamma_4$ transition.
However, the $\Gamma_3\longrightarrow\Gamma_4$ transition is not Raman active in the E$_g$ symmetry. We can rule out this CEF transition. Table~\ref{tab1} provides all Raman active transitions in the E$_g$ symmetry both for an odd and an even number of electrons on the Uranium sites as well as the mapping of the different \textit{5f} states of the Uranium ions. It appears that none of the ground states ($\Gamma_i$=$\Gamma_1$ to $\Gamma_7$) for the Uranium can be excluded. But once it is established, the final state $\Gamma_f$ of the excitation is constrained in most cases. For example, with a \textit{5f}$^2$ configuration of the Urnaium ions, if $\Gamma_i=\Gamma_1$ then $\Gamma_f=\Gamma_5$.  We also note that, with a \textit{5f}$^2$ configuration, the state $\Gamma_5$ is always involved in the excitation either as $\Gamma_i$ or $\Gamma_f$. In Table~\ref{tab1}, all the other Raman active symmetries of the excitations (in addition to the E$_g$ symmetry activity) are presented. Here, the selection rule for the CEF excitation are given in the D$_{4h}$ symmetry which is the local point symmetry of the Uranium site at high temperature. However, upon entering the hidden order state below 17 K, a change of local point symmetry is hypothesized \cite{harima_why_2010}. Then other point groups (D$_4$, C$_{4h}$, D$_{2d}$ and D$_{2h}$) must be considered.

\begin{landscape}
\begin{table}[htpb]
\centering
\renewcommand{\arraystretch}{1.8}
\begin{tabular}{|>{\centering}m{2cm}|c|c|c|c|c|c|c|c|c|c|c|c|c|c|c|c|c|c|c| }
\hline
 & \multicolumn{12}{c|}{Simple irreducible representations} & \multicolumn{6}{c|}{Double irreducible representations} \\
 & \multicolumn{12}{c|}{\textit{(5f$^2$ configuration for U)}} & \multicolumn{6}{c|}{\textit{(5f$^1$ or 5f$^3$ configuration for U)}} \\
\hline
\multirow{5}{2cm}{\begin{center}
Mapping of the irreducible representations
\end{center}}
& \multicolumn{12}{c|}{$\Gamma_1^+$ (A$_{1g}$)} & \multicolumn{6}{c|}{$\Gamma_6^+$} \\
& \multicolumn{12}{c|}{$\Gamma_2^+$ (A$_{2g}$)} & \multicolumn{6}{c|}{$\Gamma_7^+$} \\
& \multicolumn{12}{c|}{$\Gamma_3^+$ (B$_{1g}$)} & \multicolumn{6}{c|}{ } \\
& \multicolumn{12}{c|}{$\Gamma_4^+$ (B$_{2g}$)} & \multicolumn{6}{c|}{ } \\
& \multicolumn{12}{c|}{$\Gamma_5^+$ (E$_{g}$)}  & \multicolumn{6}{c|}{ } \\
\hline
Possible E$_{g}$ CEF transitions &
 \multicolumn{3}{c|}{$\Gamma_5^+$ $\rightleftarrows$ $\Gamma_1^+$}&
 \multicolumn{3}{c|}{$\Gamma_5^+$ $\rightleftarrows$ $\Gamma_2^+$}& 
 \multicolumn{3}{c|}{$\Gamma_5^+$ $\rightleftarrows$ $\Gamma_3^+$}&
 \multicolumn{3}{c|}{$\Gamma_5^+$ $\rightleftarrows$ $\Gamma_4^+$}&
 \multicolumn{2}{>{\centering}m{1.8cm}|}{$\Gamma_6^+$ $\rightleftarrows$ $\Gamma_6^+$}&
 \multicolumn{2}{>{\centering}m{1.8cm}|}{$\Gamma_6^+$ $\rightleftarrows$ $\Gamma_7^+$}&
 \multicolumn{2}{>{\centering}m{1.8cm}|}{$\Gamma_7^+$ $\rightleftarrows$ $\Gamma_7^+$}\\
\hline
Additional symmetries of the excitation  &
 \multicolumn{3}{c|}{None} &
 \multicolumn{3}{c|}{None} & 
 \multicolumn{3}{c|}{None} &
 \multicolumn{3}{c|}{None} &
 \multicolumn{2}{c|}{$\Gamma_1^+$, $\Gamma_2^+$} &
 \multicolumn{2}{c|}{$\Gamma_3^+$, $\Gamma_4^+$} &
 \multicolumn{2}{c|}{$\Gamma_1^+$, $\Gamma_2^+$} \\
\hline
\hline
Possible A$_{2g}$ CEF transitions &
 \multicolumn{4}{>{\centering}m{2.2cm}|}{$\Gamma_1^+$ $\rightleftarrows$ $\Gamma_2^+$} &
 \multicolumn{4}{>{\centering}m{2.2cm}|}{$\Gamma_3^+$ $\rightleftarrows$ $\Gamma_4^+$} & 
 \multicolumn{4}{>{\centering}m{2.2cm}|}{$\Gamma_5^+$ $\rightleftarrows$ $\Gamma_5^+$} &
 \multicolumn{3}{>{\centering}m{3cm}|}{$\Gamma_6^+$ $\rightleftarrows$ $\Gamma_6^+$ }&
 \multicolumn{3}{>{\centering}m{3cm}|}{$\Gamma_7^+$ $\rightleftarrows$ $\Gamma_7^+$}\\
\hline
Additional symmetries of the excitation  &
 \multicolumn{4}{c|}{None} &
 \multicolumn{4}{c|}{None} & 
 \multicolumn{4}{c|}{$\Gamma_1^+$, $\Gamma_3^+$, $\Gamma_4^+$} &
 \multicolumn{3}{c|}{$\Gamma_1^+$, $\Gamma_5^+$} &
 \multicolumn{3}{c|}{$\Gamma_1^+$, $\Gamma_5^+$} \\
\hline
\end{tabular}
\caption{List of the CEF excitations localized on the Uranium atoms for even and odd number of \textit{5f} electrons configurations in the D$_{4h}$ local point group symmetry as expected in URu$_2$Si$_2$. Are given the excitations which are Raman active in the E$_{g}$ symmetry (high energy excitation) and the A$_{2g}$ symmetry (low energy excitation \cite{buhot_symmetry_2014-1}). Some of these excitations are active in other symmetries \cite{koster_properties_1963}.}
\label{tab1}
\end{table}
\end{landscape}

At present, we cannot give more precise assignment to this high energy CEF transition but it clearly constrains the CEF scheme.  The only other observation of a CEF excitation hypothetized so far in URu$_2$Si$_2$ is the low energy excitation recently reported in the A$_{2g}$ symmetry at 14~\ensuremath{~\textrm{cm}^{-1}} \cite{buhot_symmetry_2014-1}. Table~\ref{tab1} presents all the CEF transitions compatible with this A$_{2g}$ symmetry. None of these transitions are excluded by our observation in the E$_g$ symmetry.  Various CEF schemes are compatible with both observations. 
For instance, in the \textit{5f$^2$} configuration, the A$_{2g}$ excitation may arise from the transition $\Gamma_3 \longrightarrow\Gamma_4$ while the E$_{g}$ excitation comes from the transition $\Gamma_3 \longrightarrow \Gamma_5$ (see supplementary of \cite{buhot_symmetry_2014-1}). In this case, no other Raman active symmetry are expected for both transitions in the D$_{4h}$ point group. 
The CEF scheme ($\Gamma_i=\Gamma_2$ and first excited state = $\Gamma_1$) proposed by Haule et al. \cite{haule_arrested_2009} is also compatible with our observations. Furthermore, we do not exclude other kind of multipolar orders \cite{suzuki_multipole_2014}.

The temperature dependence of the E$_g$ and A$_{2g}$ peaks differ strongly. Indeed, the width of the high energy one (E$_g$) is roughly constant down to the hidden order transition. Upon entering it below 17~K, the width decreases roughly by a factor 2 and the energy increases slightly by $\sim25$ \icm. However, the error on our fit of the E$_{g}$ excitation is larger at low temperature than at high temperature because the fitting range is reduced by the opening of the Kondo gap below $\sim700$\icm~\cite{buhot__2016} (Cf. Fig. \ref{fig3} b)) and the spectrum at 15 K which have been obtained with less laser power is noisier. On the contrary, the width of the low energy excitation (A$_{2g}$) diminishes linearly from 300K to 50~K (here the quasi-elastic A$_{2g}$ peak is interpreted as a precursor of the $A_{2g}$ narrow peak of the hidden order state) and it becomes inelastic and much narrower (from 40~\ensuremath{~\textrm{cm}^{-1}} at 50~K to 1~\ensuremath{~\textrm{cm}^{-1}} at 10 K) when entering the HO state \cite{buhot_symmetry_2014-1}. Opening of a gap below $\sim$50~\ensuremath{~\textrm{cm}^{-1}} in the same A$_{2g}$ symmetry may participate to this dramatic change. Although we associate the E$_{g}$ high energy excitation to a simple CEF excitation, we leave other interpretations on the nature of low energy A$_{2g}$  one open, such as a mode originating from purely itinerant electrons.

\section{Conclusion}

We have investigated the high energy modes of \urusi~by Raman scattering. Four new peaks have been observed. Three of them, seen in the A$_{1g}$ symmetry between 300 and 900\ensuremath{~\textrm{cm}^{-1}}, have been assigned to double phonon processes consistently with the \textit{ab initio} calculations of the phonon dispersion curves. A fourth broad peak observed only in the E$_{g}$ symmetry at $\sim$~760\ensuremath{~\textrm{cm}^{-1}} is attributed to a electric field excitation. This observation constrains the crystal field scheme of the \textit{5f} electrons on the Uranium sites, while allowing for the possibility to consider various CEF scheme, notably ones which are consistent with the interpretation of the low energy A$_{2g}$ excitation as to be due to a CEF excitation.

\section*{Acknowledgments}
This work was supported by the Labex SEAM (Grant No. ANR-11-IDEX-0005-02) and by the french Agence Nationale de la Recherche (ANR PRINCESS).

\clearpage

\section*{References}

%%%%%%%%%%%%%%%%%%%%%%%
%% Elsevier bibliography styles
%%%%%%%%%%%%%%%%%%%%%%%
%% To change the style, put a % in front of the second line of the current style and
%% remove the % from the second line of the style you would like to use.
%%%%%%%%%%%%%%%%%%%%%%%

%% Numbered
%\bibliographystyle{model1-num-names}

%% Numbered without titles
\bibliographystyle{model1a-num-names}

%% Harvard
%\bibliographystyle{model2-names.bst}\biboptions{authoryear}

%% Vancouver numbered
%\usepackage{numcompress}\bibliographystyle{model3-num-names}

%% Vancouver name/year
%\usepackage{numcompress}\bibliographystyle{model4-names}\biboptions{authoryear}

%% APA style
%\bibliographystyle{model5-names}\biboptions{authoryear}

%% AMA style
%\usepackage{numcompress}\bibliographystyle{model6-num-names}

%% `Elsevier LaTeX' style
%\bibliographystyle{elsarticle-num}
%%%%%%%%%%%%%%%%%%%%%%%

\bibliography{biblio}

\begin{thebibliography}{35}
\expandafter\ifx\csname natexlab\endcsname\relax\def\natexlab#1{#1}\fi
\providecommand{\url}[1]{\texttt{#1}}
\providecommand{\href}[2]{#2}
\providecommand{\path}[1]{#1}
\providecommand{\DOIprefix}{doi:}
\providecommand{\ArXivprefix}{arXiv:}
\providecommand{\URLprefix}{URL: }
\providecommand{\Pubmedprefix}{pmid:}
\providecommand{\doi}[1]{\href{http://dx.doi.org/#1}{\path{#1}}}
\providecommand{\Pubmed}[1]{\href{pmid:#1}{\path{#1}}}
\providecommand{\bibinfo}[2]{#2}
\ifx\xfnm\relax \def\xfnm[#1]{\unskip,\space#1}\fi
%Type = Article
\bibitem[{Mydosh and Oppeneer(2011)}]{mydosh_colloquium:_2011}
\bibinfo{author}{J.~A. Mydosh}, \bibinfo{author}{P.~M. Oppeneer},
  \bibinfo{journal}{Reviews of Modern Physics} \bibinfo{volume}{83}
  (\bibinfo{year}{2011}) \bibinfo{pages}{1301--1322}. \URLprefix
  \url{http://link.aps.org/doi/10.1103/RevModPhys.83.1301}.
  \DOIprefix\doi{10.1103/RevModPhys.83.1301}.
%Type = Article
\bibitem[{Mydosh and Oppeneer(2014)}]{mydosh_hidden_2014}
\bibinfo{author}{J.~Mydosh}, \bibinfo{author}{P.~Oppeneer},
  \bibinfo{journal}{Philosophical Magazine}  (\bibinfo{year}{2014}). \URLprefix
  \url{http://www.tandfonline.com/doi/abs/10.1080/14786435.2014.916428}.
%Type = Article
\bibitem[{Fujimori et~al.(2012)Fujimori, Ohkochi, Kawasaki, Yasui, Takeda,
  Okane, Saitoh, Fujimori, Yamagami, Haga, Yamamoto, Tokiwa, Ikeda, Sugai,
  Ohkuni, Kimura, and Onuki}]{fujimori_electronic_2012}
\bibinfo{author}{S.-i. Fujimori}, \bibinfo{author}{T.~Ohkochi},
  \bibinfo{author}{I.~Kawasaki}, \bibinfo{author}{A.~Yasui},
  \bibinfo{author}{Y.~Takeda}, \bibinfo{author}{T.~Okane},
  \bibinfo{author}{Y.~Saitoh}, \bibinfo{author}{A.~Fujimori},
  \bibinfo{author}{H.~Yamagami}, \bibinfo{author}{Y.~Haga},
  \bibinfo{author}{E.~Yamamoto}, \bibinfo{author}{Y.~Tokiwa},
  \bibinfo{author}{S.~Ikeda}, \bibinfo{author}{T.~Sugai},
  \bibinfo{author}{H.~Ohkuni}, \bibinfo{author}{N.~Kimura},
  \bibinfo{author}{Y.~Onuki}, \bibinfo{journal}{Journal of the Physical Society
  of Japan} \bibinfo{volume}{81} (\bibinfo{year}{2012})
  \bibinfo{pages}{014703}. \URLprefix
  \url{http://jpsj.ipap.jp/link?JPSJ/81/014703/}.
  \DOIprefix\doi{10.1143/JPSJ.81.014703}.
%Type = Article
\bibitem[{Hassinger et~al.(2008)Hassinger, Derr, Levallois, Aoki, Behnia,
  Bourdarot, Knebel, Proust, and Flouquet}]{hassinger_skutterudite_2008}
\bibinfo{author}{E.~Hassinger}, \bibinfo{author}{J.~Derr},
  \bibinfo{author}{J.~Levallois}, \bibinfo{author}{D.~Aoki},
  \bibinfo{author}{K.~Behnia}, \bibinfo{author}{F.~Bourdarot},
  \bibinfo{author}{G.~Knebel}, \bibinfo{author}{C.~Proust},
  \bibinfo{author}{J.~Flouquet}, \bibinfo{journal}{Journal of the Physical
  Society of Japan} \bibinfo{volume}{77} (\bibinfo{year}{2008})
  \bibinfo{pages}{172--179}. \URLprefix
  \url{http://journals.jps.jp/doi/abs/10.1143/JPSJS.77SA.172}.
  \DOIprefix\doi{10.1143/JPSJS.77SA.172}.
%Type = Article
\bibitem[{Booth et~al.(2016)Booth, Medling, Tobin, Baumbach, Bauer, Sokaras,
  Nordlund, and Weng}]{booth_probing_2016}
\bibinfo{author}{C.~H. Booth}, \bibinfo{author}{S.~A. Medling},
  \bibinfo{author}{J.~G. Tobin}, \bibinfo{author}{R.~E. Baumbach},
  \bibinfo{author}{E.~D. Bauer}, \bibinfo{author}{D.~Sokaras},
  \bibinfo{author}{D.~Nordlund}, \bibinfo{author}{T.-C. Weng},
  \bibinfo{journal}{Physical Review B} \bibinfo{volume}{94}
  (\bibinfo{year}{2016}) \bibinfo{pages}{045121}. \URLprefix
  \url{http://link.aps.org/doi/10.1103/PhysRevB.94.045121}.
  \DOIprefix\doi{10.1103/PhysRevB.94.045121}.
%Type = Article
\bibitem[{Kusunose and Harima(2011)}]{kusunose_hidden_2011-1}
\bibinfo{author}{H.~Kusunose}, \bibinfo{author}{H.~Harima},
  \bibinfo{journal}{Journal of the Physical Society of Japan}
  \bibinfo{volume}{80} (\bibinfo{year}{2011}) \bibinfo{pages}{084702}.
  \URLprefix \url{http://jpsj.ipap.jp/link?JPSJ/80/084702/}.
  \DOIprefix\doi{10.1143/JPSJ.80.084702}.
%Type = Article
\bibitem[{Ressouche et~al.(2012)Ressouche, Ballou, Bourdarot, Aoki, Simonet,
  Fernandez-Diaz, Stunault, and Flouquet}]{ressouche_hidden_2012}
\bibinfo{author}{E.~Ressouche}, \bibinfo{author}{R.~Ballou},
  \bibinfo{author}{F.~Bourdarot}, \bibinfo{author}{D.~Aoki},
  \bibinfo{author}{V.~Simonet}, \bibinfo{author}{M.~T. Fernandez-Diaz},
  \bibinfo{author}{A.~Stunault}, \bibinfo{author}{J.~Flouquet},
  \bibinfo{journal}{Physical Review Letters} \bibinfo{volume}{109}
  (\bibinfo{year}{2012}) \bibinfo{pages}{067202}. \URLprefix
  \url{http://link.aps.org/doi/10.1103/PhysRevLett.109.067202}.
  \DOIprefix\doi{10.1103/PhysRevLett.109.067202}.
%Type = Article
\bibitem[{Haule and Kotliar(2009)}]{haule_arrested_2009}
\bibinfo{author}{K.~Haule}, \bibinfo{author}{G.~Kotliar},
  \bibinfo{journal}{Nature Physics} \bibinfo{volume}{5} (\bibinfo{year}{2009})
  \bibinfo{pages}{796--799}. \URLprefix
  \url{http://www.nature.com/doifinder/10.1038/nphys1392}.
  \DOIprefix\doi{10.1038/nphys1392}.
%Type = Article
\bibitem[{Ikeda et~al.(2012)Ikeda, Suzuki, Arita, Takimoto, Shibauchi, and
  Matsuda}]{ikeda_emergent_2012}
\bibinfo{author}{H.~Ikeda}, \bibinfo{author}{M.-T. Suzuki},
  \bibinfo{author}{R.~Arita}, \bibinfo{author}{T.~Takimoto},
  \bibinfo{author}{T.~Shibauchi}, \bibinfo{author}{Y.~Matsuda},
  \bibinfo{journal}{Nature Physics} \bibinfo{volume}{8} (\bibinfo{year}{2012})
  \bibinfo{pages}{528--533}. \URLprefix
  \url{http://www.nature.com/doifinder/10.1038/nphys2330}.
  \DOIprefix\doi{10.1038/nphys2330}.
%Type = Article
\bibitem[{Rau and Kee(2012)}]{rau_hidden_2012}
\bibinfo{author}{J.~G. Rau}, \bibinfo{author}{H.-Y. Kee},
  \bibinfo{journal}{Physical Review B} \bibinfo{volume}{85}
  (\bibinfo{year}{2012}) \bibinfo{pages}{245112}. \URLprefix
  \url{http://link.aps.org/doi/10.1103/PhysRevB.85.245112}.
  \DOIprefix\doi{10.1103/PhysRevB.85.245112}.
%Type = Article
\bibitem[{Suzuki and Ikeda(2014)}]{suzuki_multipole_2014}
\bibinfo{author}{M.-T. Suzuki}, \bibinfo{author}{H.~Ikeda},
  \bibinfo{journal}{Physical Review B} \bibinfo{volume}{90}
  (\bibinfo{year}{2014}) \bibinfo{pages}{184407}. \URLprefix
  \url{http://link.aps.org/doi/10.1103/PhysRevB.90.184407}.
  \DOIprefix\doi{10.1103/PhysRevB.90.184407}.
%Type = Article
\bibitem[{Chandra et~al.(2002)Chandra, Coleman, Mydosh, and
  Tripathi}]{chandra_hidden_2002}
\bibinfo{author}{P.~Chandra}, \bibinfo{author}{P.~Coleman},
  \bibinfo{author}{J.~A. Mydosh}, \bibinfo{author}{V.~Tripathi},
  \bibinfo{journal}{Nature} \bibinfo{volume}{417} (\bibinfo{year}{2002})
  \bibinfo{pages}{831--834}. \URLprefix
  \url{http://www.nature.com/nature/journal/v417/n6891/abs/nature00795.html}.
  \DOIprefix\doi{10.1038/nature00795}.
%Type = Article
\bibitem[{Fujimoto(2011)}]{fujimoto_spin_2011}
\bibinfo{author}{S.~Fujimoto}, \bibinfo{journal}{Physical Review Letters}
  \bibinfo{volume}{106} (\bibinfo{year}{2011}) \bibinfo{pages}{196407}.
  \URLprefix \url{http://link.aps.org/doi/10.1103/PhysRevLett.106.196407}.
%Type = Article
\bibitem[{Riseborough et~al.(2012)Riseborough, Coqblin, and
  Magalhães}]{riseborough_phase_2012}
\bibinfo{author}{P.~S. Riseborough}, \bibinfo{author}{B.~Coqblin},
  \bibinfo{author}{S.~G. Magalhães}, \bibinfo{journal}{Physical Review B}
  \bibinfo{volume}{85} (\bibinfo{year}{2012}) \bibinfo{pages}{165116}.
  \URLprefix \url{http://link.aps.org/doi/10.1103/PhysRevB.85.165116}.
  \DOIprefix\doi{10.1103/PhysRevB.85.165116}.
%Type = Article
\bibitem[{Das(2014)}]{das_imprints_2014}
\bibinfo{author}{T.~Das}, \bibinfo{journal}{Physical Review B}
  \bibinfo{volume}{89} (\bibinfo{year}{2014}) \bibinfo{pages}{045135}.
  \URLprefix \url{http://link.aps.org/doi/10.1103/PhysRevB.89.045135}.
  \DOIprefix\doi{10.1103/PhysRevB.89.045135}.
%Type = Article
\bibitem[{Pépin et~al.(2011)Pépin, Norman, Burdin, and
  Ferraz}]{pepin_modulated_2011}
\bibinfo{author}{C.~Pépin}, \bibinfo{author}{M.~R. Norman},
  \bibinfo{author}{S.~Burdin}, \bibinfo{author}{A.~Ferraz},
  \bibinfo{journal}{Physical Review Letters} \bibinfo{volume}{106}
  (\bibinfo{year}{2011}) \bibinfo{pages}{106601}. \URLprefix
  \url{http://link.aps.org/doi/10.1103/PhysRevLett.106.106601}.
%Type = Article
\bibitem[{Thomas et~al.(2013)Thomas, Burdin, Pépin, and
  Ferraz}]{thomas_three-dimensional_2013}
\bibinfo{author}{C.~Thomas}, \bibinfo{author}{S.~Burdin},
  \bibinfo{author}{C.~Pépin}, \bibinfo{author}{A.~Ferraz},
  \bibinfo{journal}{Physical Review B} \bibinfo{volume}{87}
  (\bibinfo{year}{2013}) \bibinfo{pages}{014422}. \URLprefix
  \url{http://link.aps.org/doi/10.1103/PhysRevB.87.014422}.
  \DOIprefix\doi{10.1103/PhysRevB.87.014422}.
%Type = Article
\bibitem[{Elgazzar et~al.(2009)Elgazzar, Rusz, Amft, Oppeneer, and
  Mydosh}]{elgazzar_hidden_2009}
\bibinfo{author}{S.~Elgazzar}, \bibinfo{author}{J.~Rusz},
  \bibinfo{author}{M.~Amft}, \bibinfo{author}{P.~M. Oppeneer},
  \bibinfo{author}{J.~A. Mydosh}, \bibinfo{journal}{Nature Materials}
  \bibinfo{volume}{8} (\bibinfo{year}{2009}) \bibinfo{pages}{337--341}.
  \URLprefix \url{http://www.nature.com/doifinder/10.1038/nmat2395}.
  \DOIprefix\doi{10.1038/nmat2395}.
%Type = Article
\bibitem[{Chandra et~al.(2015)Chandra, Coleman, and
  Flint}]{chandra_hastatic_2015-1}
\bibinfo{author}{P.~Chandra}, \bibinfo{author}{P.~Coleman},
  \bibinfo{author}{R.~Flint}, \bibinfo{journal}{Physical Review B}
  \bibinfo{volume}{91} (\bibinfo{year}{2015}) \bibinfo{pages}{205103}.
  \URLprefix \url{http://link.aps.org/doi/10.1103/PhysRevB.91.205103}.
  \DOIprefix\doi{10.1103/PhysRevB.91.205103}.
%Type = Article
\bibitem[{Buhot et~al.(2014)Buhot, Méasson, Gallais, Cazayous, Sacuto,
  Lapertot, and Aoki}]{buhot_symmetry_2014-1}
\bibinfo{author}{J.~Buhot}, \bibinfo{author}{M.-A. Méasson},
  \bibinfo{author}{Y.~Gallais}, \bibinfo{author}{M.~Cazayous},
  \bibinfo{author}{A.~Sacuto}, \bibinfo{author}{G.~Lapertot},
  \bibinfo{author}{D.~Aoki}, \bibinfo{journal}{Physical Review Letters}
  \bibinfo{volume}{113} (\bibinfo{year}{2014}) \bibinfo{pages}{266405}.
  \URLprefix \url{http://link.aps.org/doi/10.1103/PhysRevLett.113.266405}.
  \DOIprefix\doi{10.1103/PhysRevLett.113.266405}.
%Type = Article
\bibitem[{Kung et~al.(2015)Kung, Baumbach, Bauer, Thorsmølle, Zhang, Haule,
  Mydosh, and Blumberg}]{kung_chirality_2015}
\bibinfo{author}{H.-H. Kung}, \bibinfo{author}{R.~E. Baumbach},
  \bibinfo{author}{E.~D. Bauer}, \bibinfo{author}{V.~K. Thorsmølle},
  \bibinfo{author}{W.-L. Zhang}, \bibinfo{author}{K.~Haule},
  \bibinfo{author}{J.~A. Mydosh}, \bibinfo{author}{G.~Blumberg},
  \bibinfo{journal}{Science}  (\bibinfo{year}{2015}) \bibinfo{pages}{1259729}.
  \URLprefix
  \url{http://www.sciencemag.org/content/early/2015/02/12/science.1259729}.
  \DOIprefix\doi{10.1126/science.1259729}.
%Type = Article
\bibitem[{Devereaux and Hackl(2007)}]{devereaux_inelastic_2007}
\bibinfo{author}{T.~P. Devereaux}, \bibinfo{author}{R.~Hackl},
  \bibinfo{journal}{Review of Modern Physics} \bibinfo{volume}{79}
  (\bibinfo{year}{2007}) \bibinfo{pages}{175--233}. \URLprefix
  \url{http://link.aps.org/doi/10.1103/RevModPhys.79.175}.
  \DOIprefix\doi{10.1103/RevModPhys.79.175}.
%Type = Book
\bibitem[{Cardona and Güntherodt(2000)}]{cardona_light_2000}
\bibinfo{author}{M.~Cardona}, \bibinfo{author}{G.~Güntherodt},
  \bibinfo{title}{Light {Scattering} in {Solids} {VII}: {Crystal}-field and
  magnetic excitations.}, \bibinfo{publisher}{Springer}, \bibinfo{year}{2000}.
%Type = Article
\bibitem[{Ogita et~al.(2009)Ogita, Kojima, Hasegawa, Udagawa, Sugawara, and
  Sato}]{ogita_crystal_2009}
\bibinfo{author}{N.~Ogita}, \bibinfo{author}{R.~Kojima},
  \bibinfo{author}{T.~Hasegawa}, \bibinfo{author}{M.~Udagawa},
  \bibinfo{author}{H.~Sugawara}, \bibinfo{author}{H.~Sato},
  \bibinfo{journal}{Journal of Physics: Conference Series}
  \bibinfo{volume}{150} (\bibinfo{year}{2009}) \bibinfo{pages}{042147}.
  \URLprefix
  \url{http://stacks.iop.org/1742-6596/150/i=4/a=042147?key=crossref.3998ba143a0b8e20cc61f1c13a7ddf16}.
  \DOIprefix\doi{10.1088/1742-6596/150/4/042147}.
%Type = Article
\bibitem[{Buhot et~al.(2015)Buhot, Méasson, Gallais, Cazayous, Sacuto,
  Bourdarot, Raymond, Lapertot, Aoki, Regnault, Ivanov, Piekarz, Parlinski,
  Legut, Homes, Lejay, and Lobo}]{buhot_lattice_2015}
\bibinfo{author}{J.~Buhot}, \bibinfo{author}{M.~A. Méasson},
  \bibinfo{author}{Y.~Gallais}, \bibinfo{author}{M.~Cazayous},
  \bibinfo{author}{A.~Sacuto}, \bibinfo{author}{F.~Bourdarot},
  \bibinfo{author}{S.~Raymond}, \bibinfo{author}{G.~Lapertot},
  \bibinfo{author}{D.~Aoki}, \bibinfo{author}{L.~P. Regnault},
  \bibinfo{author}{A.~Ivanov}, \bibinfo{author}{P.~Piekarz},
  \bibinfo{author}{K.~Parlinski}, \bibinfo{author}{D.~Legut},
  \bibinfo{author}{C.~C. Homes}, \bibinfo{author}{P.~Lejay},
  \bibinfo{author}{R.~P. S.~M. Lobo}, \bibinfo{journal}{Physical Review B}
  \bibinfo{volume}{91} (\bibinfo{year}{2015}) \bibinfo{pages}{035129}.
  \URLprefix \url{http://link.aps.org/doi/10.1103/PhysRevB.91.035129}.
  \DOIprefix\doi{10.1103/PhysRevB.91.035129}.
%Type = Article
\bibitem[{Aoki et~al.(2010)Aoki, Bourdarot, Hassinger, Knebel, Miyake, Raymond,
  Taufour, and Flouquet}]{aoki_field_2010}
\bibinfo{author}{D.~Aoki}, \bibinfo{author}{F.~Bourdarot},
  \bibinfo{author}{E.~Hassinger}, \bibinfo{author}{G.~Knebel},
  \bibinfo{author}{A.~Miyake}, \bibinfo{author}{S.~Raymond},
  \bibinfo{author}{V.~Taufour}, \bibinfo{author}{J.~Flouquet},
  \bibinfo{journal}{Journal of Physics: Condensed Matter} \bibinfo{volume}{22}
  (\bibinfo{year}{2010}) \bibinfo{pages}{164205}. \URLprefix
  \url{http://stacks.iop.org/0953-8984/22/i=16/a=164205?key=crossref.ebb4de9a5c8d62596981c192f5734524}.
  \DOIprefix\doi{10.1088/0953-8984/22/16/164205}.
%Type = Article
\bibitem[{Buhot et~al.(2013)Buhot, Méasson, Gallais, Cazayous, Sacuto,
  Lapertot, and Aoki}]{buhot_raman_2013}
\bibinfo{author}{J.~Buhot}, \bibinfo{author}{M.-A. Méasson},
  \bibinfo{author}{Y.~Gallais}, \bibinfo{author}{M.~Cazayous},
  \bibinfo{author}{A.~Sacuto}, \bibinfo{author}{G.~Lapertot},
  \bibinfo{author}{D.~Aoki}, \bibinfo{journal}{Journal of the Korean Physical
  Society} \bibinfo{volume}{62} (\bibinfo{year}{2013})
  \bibinfo{pages}{1427--1430}. \URLprefix
  \url{http://dx.doi.org/10.3938/jkps.62.1427}.
%Type = Book
\bibitem[{Hayes and Loudon(2004)}]{hayes_scattering_2004}
\bibinfo{author}{W.~Hayes}, \bibinfo{author}{R.~Loudon},
  \bibinfo{title}{Scattering of {Light} by {Crystals}}, Dover {Books} on
  {Physics}, \bibinfo{publisher}{Dover Publications}, \bibinfo{year}{2004}.
  \URLprefix \url{http://books.google.fr/books?id=8N4rU_gtHgAC}.
%Type = Article
\bibitem[{Kresse and Furthmüller(1996)}]{kresse_efficient_1996-1}
\bibinfo{author}{G.~Kresse}, \bibinfo{author}{J.~Furthmüller},
  \bibinfo{journal}{Physical Review B} \bibinfo{volume}{54}
  (\bibinfo{year}{1996}) \bibinfo{pages}{11169--11186}. \URLprefix
  \url{http://link.aps.org/doi/10.1103/PhysRevB.54.11169}.
%Type = Article
\bibitem[{Parlinski et~al.(1997)Parlinski, Li, and
  Kawazoe}]{parlinski_first-principles_1997}
\bibinfo{author}{K.~Parlinski}, \bibinfo{author}{Z.~Q. Li},
  \bibinfo{author}{Y.~Kawazoe}, \bibinfo{journal}{Physical Review Letters}
  \bibinfo{volume}{78} (\bibinfo{year}{1997}) \bibinfo{pages}{4063--4066}.
  \URLprefix \url{http://link.aps.org/doi/10.1103/PhysRevLett.78.4063}.
  \DOIprefix\doi{10.1103/PhysRevLett.78.4063}.
%Type = Article
\bibitem[{Park et~al.(2002)Park, McEwen, and Bull}]{park_high-energy_2002}
\bibinfo{author}{J.-G. Park}, \bibinfo{author}{K.~McEwen},
  \bibinfo{author}{M.~Bull}, \bibinfo{journal}{Physical Review B}
  \bibinfo{volume}{66} (\bibinfo{year}{2002}). \URLprefix
  \url{http://link.aps.org/doi/10.1103/PhysRevB.66.094502}.
  \DOIprefix\doi{10.1103/PhysRevB.66.094502}.
%Type = Article
\bibitem[{Santini and Amoretti(1994)}]{santini_crystal_1994}
\bibinfo{author}{P.~Santini}, \bibinfo{author}{G.~Amoretti},
  \bibinfo{journal}{Physical Review Letters} \bibinfo{volume}{73}
  (\bibinfo{year}{1994}) \bibinfo{pages}{1027}. \URLprefix
  \url{http://prl.aps.org/abstract/PRL/v73/i7/p1027_1}.
%Type = Article
\bibitem[{Harima et~al.(2010)Harima, Miyake, and Flouquet}]{harima_why_2010}
\bibinfo{author}{H.~Harima}, \bibinfo{author}{K.~Miyake},
  \bibinfo{author}{J.~Flouquet}, \bibinfo{journal}{Journal of the Physical
  Society of Japan} \bibinfo{volume}{79} (\bibinfo{year}{2010})
  \bibinfo{pages}{033705}. \URLprefix
  \url{http://jpsj.ipap.jp/link?JPSJ/79/033705/}.
  \DOIprefix\doi{10.1143/JPSJ.79.033705}.
%Type = Article
\bibitem[{Buhot(2016)}]{buhot__2016}
\bibinfo{author}{J.~Buhot}, \bibinfo{journal}{et al., in preparation}
  (\bibinfo{year}{2016}).
%Type = Book
\bibitem[{Koster(1963)}]{koster_properties_1963}
\bibinfo{author}{G.~F. Koster}, \bibinfo{title}{Properties of the thirty-two
  point groups}, \bibinfo{publisher}{M.I.T. Press},
  \bibinfo{address}{Cambridge}, \bibinfo{year}{1963}.

\end{thebibliography}

\end{document}